\documentstyle[epsf,prl,twocolumn,aps,floats]{revtex} 
\begin{document}
\topmargin = -0.6cm
\overfullrule 0pt

\twocolumn[\hsize\textwidth\columnwidth\hsize\csname
@twocolumnfalse\endcsname

\title{
\hfill{\small IFT-P.057/99} \\
 \hfill{\small hep-ph/XXXXXX}\\
\vglue 0.5cm
Flavor changing models with strictly massless neutrinos}
\author{ 
M.\ M.\ Guzzo$^{1}$, 
H.\ Nunokawa$^{1}$, \\
O.\ L.\ G.\ Peres$^{1}$, 
V. Pleitez$^{2}$,  and 
R.\ Zukanovich Funchal$^{3}$ }
\address{\sl 
$^1$ Instituto de F\' {\i}sica Gleb Wataghin\\
    Universidade Estadual de Campinas, UNICAMP\\    
    13083-970 -- Campinas, Brazil \\
$^2$ Instituto de F\' {\i}sica Te\'orica \\
    Universidade Estadual Paulista\\
    R. Pamplona 145, 01405-900 S\~ao Paulo, Brazil \\
$^3$ Instituto de F\'{\i}sica \\ 
       Universidade de S\~ao Paulo \\
    C.\ P.\ 66.318, 05389-970 S\~ao Paulo, Brazil. }
\maketitle
\vspace{.5cm}

\hfuzz=25pt
\begin{abstract} 
Theoretical models which present flavor changing neutrino interactions
and simultaneously prevent this particle from acquiring any mass exist.
We discuss some of them and their predictions for neutrino oscillations
in matter which can account for the solar neutrino anomaly as well as
the zenith-angle dependence of the deficit of atmospheric neutrinos
observed by the SuperKamiokande experiment without invoking, therefore,
neutrino masses nor mixing.
\end{abstract}

\vskip2pc]

\section{Introduction}
\label{sec:intro}
Flavor changing neutrino interactions (FCNI) can induce neutrino
oscillations in matter. This phenomenon was first investigated by 
Wolfenstein~\cite{lw1} who pointed out that interactions in a medium modify 
the dispersion relations of particles traveling through. Wolfenstein effect 
generates quantum phases in time evolution of phenomenological neutrinos 
eigenstates which consequently can oscillate. These oscillations can be 
resonantly enhanced even if neutrinos are massless and no mixing in the vacuum
exists~\cite{gmp,valle87}.  Similar phenomenon happens when neutrino masses are
introduced and mixing angles in matter are induced by flavor changing
interactions~\cite{gmp,er}. Wolfenstein effect has
been  invoked to obtain a good fit of the solar neutrino
observations~\cite{solarexp}  when FCNI is assumed in the massless
neutrino context~\cite{gmp,kb97} or when the  neutrinos are assumed
massive~\cite{gmp,er,FC-mass}.  Furthermore it was recently pointed out that
the SuperKamiokande results~\cite{sk} showing a strong zenith-angle
dependence of the  $\mu$ flux induced by atmospheric $\mu$-neutrinos, 
which have been interpreted as an indication of neutrino
oscillations~\cite{sk1,ourwork},  can also be understood assuming
non-resonant FCNI with ordinary matter in the Earth~\cite{rpatm,comment}.

The presence of flavor changing neutrino-matter interactions implies a
non-trivial structure for the neutrino evolution Hamiltonian in matter
even if massless neutrinos and no mixing in the vacuum is assumed.  The
evolution equations describing  the $\nu_\alpha \Rightarrow \nu_\beta$
transitions ($\alpha,\beta = e, \mu,\tau$ are flavor indices) are given
by~\cite{gmp,valle87}:
\begin{eqnarray} &  i{\displaystyle{d \Psi_{\alpha\beta}}\over 
\displaystyle{dr}} 
\nonumber
= \\ & \!\!\!\!\hskip .3cm  \hskip-.1cm
\sqrt{2}\,G_F \left( \begin{array}{cc} 0 &  \epsilon^f_{\alpha\beta} n_f(r)
\\ \epsilon^f_{\alpha\beta} n_f(r)& \epsilon_{\alpha\beta} '^f n_f(r) -
\delta_{e\alpha} 
n_e(r)\end{array} \right)
\Psi_{\alpha\beta},
\label{motion} 
\end{eqnarray}
where $\Psi_{\alpha\beta}=(\nu_\alpha~~ \nu_\beta)^T$;  
$\nu_\alpha \equiv \nu_\alpha (r)$ and $\nu_\beta\equiv \nu_\beta
(r)$, are the probability amplitudes to find these neutrinos at a
distance $r$ from their creation position, $\sqrt{2}\,G_F n_f(r)
\epsilon^f_{\alpha\beta}$ is the flavor-changing $\nu_\alpha + f \to \nu_\beta 
+f$ forward scattering amplitude with the interacting fermion $f$ (charged
lepton, $d$-like quark or $u$-like quark) and $\sqrt{2}\,G_F n_f(r)
\epsilon_{\alpha\beta} '^f$ is the difference between the flavor diagonal
$\nu_\alpha - f$ and $\nu_\beta - f$ elastic forward scattering
amplitudes, with $n_f(r)$ being the number density of the fermions
which induce such processes. In all cases  of practical interest,
electron-neutrinos will coherently scatter off the electrons present in
matter through   standard electroweak charged currents which introduce
non trivial contributions to the neutrino evolution equations. These
contributions are taken into account by the term  $\sqrt 2 G_F
\delta_{e\alpha} n_e(r)$ in Eq.~(\ref{motion}). Therefore, when the
electron neutrino evolves in matter ($\alpha = e$) and the fermion 
which the neutrino scatter off is not electron ($f \neq  e$), a resonance can 
occur if the condition $\epsilon_{e\beta} '^f n_f(r) = n_e(r)$ is fulfilled at 
least in one layer of the matter crossed by the neutrinos~\cite{gmp}. In the case
where no electron-neutrino participates in the oscillation, no resonance
can happen.

In any case, flavor oscillations induced by FCNI can be relevant for
understanding a considerable range of experimental data involving
neutrinos which cannot be fitted by using only standard electroweak
inputs. The possibility that these oscillations occur even in the absence
of neutrino masses and mixing open a new perspective for
interpretation of current neutrino data.  Our purpose in this paper is
to investigate how feasible this picture can be in several models which
allow FCNI.

\section{FCNI: A general review }
\label{sec:fcni}

Although FCNI do not appear in the standard model of the electroweak
interactions they are quite naturally present in many of  its
extensions. Most of them, however, do not naturally admit strictly
massless neutrinos. An example of these models are  the $R$-parity
broken supersymmetric models~\cite{r-parity} which were invoked in the
relevant analyses for the solar and atmospheric neutrinos previously
cited~\cite{gmp,er,rpatm}. In these models, the same
FCNI interactions that induce the $\epsilon^f_{\alpha\beta}$ and 
$\epsilon'^f_{\alpha\beta}$ contributions to the Hamiltonian in 
Eq.~(\ref{motion}) induce also contributions to the neutrino mass at the 
1-loop level.  A fine tuning in the model parameters is needed in order 
to keep neutrino masses negligible while endowing 
$\epsilon^f_{\alpha\beta}$ and $\epsilon'^f_{\alpha\beta}$ with appropriate
values to fit the solar and atmospheric neutrino anomalies.

There are extensions of the standard $SU(2)\otimes U(1)$
model or of its $SU(5)$ grand unification extension in which neutrinos get 
naturally a small Dirac mass~\cite{rw}. 
In order to obtain such effect extra symmetry and fields are added to the 
minimal models. For instance, in the $SU(2)\otimes U(1)$ model besides the 
right-handed neutrino $\nu_R$, a pair of extra singlets $s_L$ and $s_R$ and 
one scalar singlet are 
added. The extra symmetry imply that for each family there are one light and 
one heavy neutral leptons. 
In this case the FCNI needed to induce non-zero 
$\epsilon^f_{\alpha\beta}$ and $\epsilon'^f_{\alpha\beta}$ are mediated by the
gauge vector boson $W$ and
for this reason the only free parameters are the mixing angles. 
This has to be confronted with 
constraints imposed by the experimental limits on lepton number violating 
decays such as  
$\mu\to e\gamma$ in a similar way we will do in Sec.~\ref{sec:cons}

Also in this context  it is worth to mention other models 
like the left-right
symmetric one and their SO(10) grand unified extensions~\cite{ww}. 
In this case three of the neutrinos
can be massless at arbitrary order in perturbation theory. However,
the size of 
flavor changing neutral currents in the neutrino sector
depends on the value of the mass scale related to the pattern of symmetry 
breaking that reduces $SO(10)$ to $SU(3)_c\otimes SU(2)_L\otimes U(1)$.
Only if $SO(10)$ breaks first to $SU(2)_L\otimes SU(2)_R\otimes U(1)$ it is
possible to have intermediated mass scale related to the energy at which 
$SU(2)_R$ is broken. In general, grand-unified theories present both 
massless neutrinos and
flavor changing interactions~\cite{gg}. Nevertheless the effective flavor
changing neutral currents are
negligibly small since they are inversely proportional to the squared
mass of the exotic vector boson which is fixed by the condition of
unification.  This is the case of all non-supersymmetric grand-unified
models since in these models there exist the hierarchy problem.

A version of the Zee's model~\cite{zee} in which only one extra singly
charged scalar singlet  (besides the usual doublet) and right-handed
neutrinos are added  was considered in Ref.~\cite{fy88}. In the
simplest case the constraint coming from the
muon decay implies that the atmospheric neutrino flux is reduced 
at most  20\%. A way to overcome this difficulty is to  add
another singly charged scalar singlet. In this case the neutrino masses
are also arbitrary 
and the constraints of the muon and other 
leptonic decays can be evaded. Notice that if we do not add
right-handed neutrinos FCNI take place only for anti-neutrinos.

Nevertheless there are models where the mechanism pointed out in
Refs.~\cite{gmp} and \cite{rpatm} can be exactly realized, keeping neutrinos 
strictly massless, while FCNI can exist with the required 
intensity. 
Basically, vanishing neutrino masses are guaranteed in these models due to 
the conservation of the total leptonic number $L$. Then, by imposing that 
there are no right-handed neutrinos to avoid Dirac masses, 
no vacuum expectation values 
associated with neutral scalars to prevent spontaneous violation of $L$ and, 
consequently,  no majoron-like Goldstone boson is generated, neutrino masses 
can be kept vanishing. Obviously, no $L$ violation in the scalar sector is 
also allowed, like as in the $R$-parity violating supersymmetric models, since 
in this case neutrino masses can be generated radiatively~\cite{fkl}.

In this paper we show that these conditions are naturally fulfilled in some 
theoretical scenarios and therefore FCNI can coexist with strictly massless 
neutrinos. In Sec.~\ref{sec:331} we consider the gauge model  based on 
$SU(3)_c\otimes SU(3)_L\otimes U(1)_N$ symmetry ( 331 model)~\cite{331} and
in Sec.~\ref{sec:mhesm} we consider the multi-Higgs doublet extension of the
standard model~\cite{higgshunter}. 

\section{FCNI in 331 models}
\label{sec:331}

Although the standard model of electroweak interactions accommodates all the
present experimental results, it is not able to give answer for
some questions in particle physics. One of these questions is the 
family replication problem, which can have an elegant solution in the
simplest chiral extension of standard model, the gauge model  based on 
$SU(3)_c\otimes SU(3)_L\otimes U(1)_N$ symmetry which broken to 
$SU(3)_c\otimes SU(2)_L\otimes U(1)_N$ in some energy scale higher
than the Fermi one. In the 331 model of Ref.~\cite{331} neutrinos are kept 
massless as long as no extra  right-handed neutrinos are added to the particle
content of the model and the total lepton number $L$ is
conserved in the full Lagrangian~\cite{fkl}. 
We will be assuming here that this is the case. 
The 331 phenomenology was studied in many different
contexts~\cite{331pheno} but for the first time it is applied  for
neutrino phenomenology.
The Yukawa interactions in the lepton sector are~\cite{331,331a}
\begin{eqnarray}
-{\cal L}_Y&=&\frac{1}{\sqrt2}\,\overline{\nu}_L\,{\cal K}_{LR}
l_RH^+_1+
\frac{1}{\sqrt2}\,\bar{l}_L\,{\cal K}_{LL} \nu^c_RH^-_2
\nonumber \\ &+&
\frac{1}{2}\,\bar{l}_L\,{\cal K}_{LL}(l_L)^cH^{--}_1 
+\frac{1}{2}\,\overline{l^c}_L\,{\cal K}_{RR}l_RH^{++}_2
\nonumber \\ \mbox{}
&+&
 2\;\overline{\nu}_L\, {\cal K}^\prime_{LR}l_R\eta^+_1
-2\;\bar{l}_L\,{\cal K}^\prime_{LL}\nu^c_R\eta^-_2
+H.c.,
\label{yu331}
\end{eqnarray}
where ${\cal K}_{LL}=E^\dagger_LGE^*_L$, ${\cal K}_{RR}=E^T_RGE_R$,
${\cal K}_{LR}=E^\dagger_LGE_R$; ${\cal K}^\prime_{LR}=
E^\dagger_L G^\prime E_R$, ${\cal K}^\prime_{LL}=E^\dagger_LG^\prime E^*_L$;
$G$ and $G'$ are symmetric and antisymmetric (they can be complex) Yukawa 
matrices. The symmetric ( antisymmetric) property of the matrices above are 
because the correspondent Higgs field is in triplet 
(sextet) representation~\cite{331,331a}. $E_{R,L}$ are the right- and 
left-handed mixing unitary matrices in the lepton sector relating symmetry 
eigenstates (primed fields) with mass eigenstates 
(unprimed fields)~\cite{liung}:
\begin{equation}
l'_L=E_Ll_L,\quad l'_R=E_Rl_R,\quad \nu'_L=E_L\nu_L,
\label{sme}
\end{equation}
where we have redefined the neutrino fields so that there is no mixing in the
current coupled to the gauge vector boson $W$. Note, therefore, that while 
${\cal K}_{LL}$ and ${\cal K}_{RR}$ are symmetric matrices and  $K'_{LL}$
is an  antisymmetric matrix,
no symmetry relation appears in ${\cal K}_{LR}$
and ${\cal K}'_{LR}$. None of the couplings in Eq.~(\ref{yu331}) depends 
directly on  the charged lepton masses and all matrices in
Eq.~(\ref{yu331}) are not unitary. When $G$ and $G'$ are real matrices the
matrices ${\cal K}_{LL}$ and ${\cal K}_{RR}$ (and the respective
primed matrices) are hermitian. All scalars in Eq.~(\ref{yu331}) are
still symmetry eigenstates. 

The interactions  of the leptons with the gauge vector boson $V$~\cite{331} 
induce also FCNI and are given by 
\begin{equation}
-{\cal L}_V= \bar{l}_R\gamma^\mu {\cal K}\nu^c_RV^-_\mu+H.c.,
\label{V}
\end{equation}
with 
${\cal K}=(g_3/\sqrt2) E^\dagger_RE_L^*$ being a unitary matrix and $g_3$ is 
the coupling  of the gauge vector boson $V$  with the leptons. This is not the 
$g$ coupling of standard model gauge vector boson $W$.    

We can now write the expressions for the parameter $\epsilon^f_{\alpha\beta}$ 
entering in 
Eq.~(\ref{motion}) and for the analogous parameter 
$\bar{\epsilon}^f_{\alpha\beta}$, when 
anti-neutrinos are evolving instead of neutrinos, in the light of this model. 
We present, in Figs. 1, 2 and 3 the diagrams which illustrate
scalar contributions for neutrinos, scalar contributions for 
anti-neutrinos and vector contributions for anti-neutrinos, 
respectively, in the 331 model.
From Figs.~\ref{fig1}, \ref{fig2} and \ref{fig3} and      
Eqs.~(\ref{yu331}) and (\ref{V}), we write the nonstandard 
contributions to $\epsilon^e_{\alpha\beta}$ and 
$\bar{\epsilon}^e_{\alpha\beta}$ (the superscript $e$ 
denotes that the relevant interactions are those with electrons):
\begin{figure}[ht]
\centering\leavevmode
\epsfxsize=150pt
\epsfbox{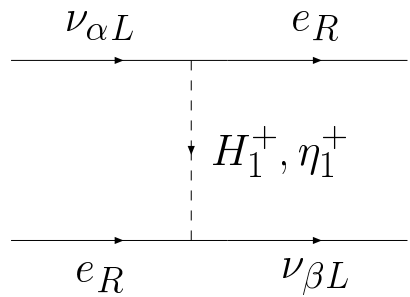}
\vglue -0.01cm
\caption{Scalar contributions to FCNI in the 331 model. }
\label{fig1}
\end{figure}

\begin{figure}[ht]
\centering\leavevmode
\epsfxsize=150pt
\epsfbox{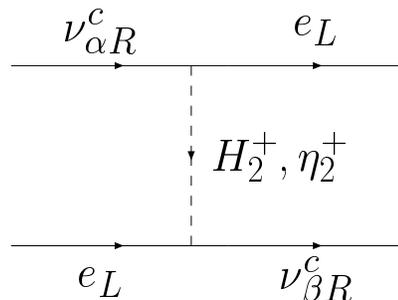}
\vglue -0.01cm
\caption{Scalar contributions to anti-neutrinos FCNI in 331 model.}
\label{fig2}
\end{figure}

\begin{mathletters}
\label{epsnunubar}
\begin{eqnarray}
\epsilon^e_{\alpha\beta}&=&\frac{1}{4\sqrt2G_FM^2_W}\;\left[
\left({\cal K}^\dagger _{LR} \right)_{1\alpha}\!\!
\left( {\cal K}_{LR}\right)_{\beta1}x\right. 
\nonumber \\ &+& \left. \left({\cal K}^{\prime\dagger}_{LR} \right)_{1\alpha }
\!\!\left({\cal K}^\prime_{LR} \right)_{\beta1}y\right],
\label{epsnu}
\end{eqnarray}
where $x=M^2_W/M^2_{H_1}$, $y=M^2_W/M^2_{\eta_1}$, and
\begin{eqnarray}
\bar{\epsilon}^e_{\alpha\beta}&=&
\frac{1}{4\sqrt2G_FM^2_W}\;\left[
\left({\cal K} \right)_{1\alpha }\!\!
\left( {\cal K}^\dagger\right)_{\beta1}z' + 
\left( {\cal K}_{LL}\right)_{1\alpha }
\!\!\left({\cal K}^\dagger_{LL} 
\right)_{\beta1}x' \right. \nonumber \\ &\!+\!&\left.
\left( {\cal K}^{\prime}_{LL}\right)_{1\alpha }
\!\!\left( {\cal K}^{\prime\dagger}_{LL}
\right)_{\beta1}y'\right],
\label{epsnubar}
\end{eqnarray}
\end{mathletters}
where $z'=M^2_W/M^2_{V}$, $x'=M^2_W/M^2_{H_2}$, $y'=M^2_W/m^2_{\eta_2}$. 
Note also that we are assuming that all the matrices entering above are real.
In case they are complex, only their real part will contribute to 
${\epsilon}^e_{\alpha\beta}$ and $\bar{\epsilon}^e_{\alpha\beta}$.
In this paper we always  assume the convention  that
Greek subscript  stands for neutrinos and  Arabic subscript for charged 
leptons. 

\begin{figure}[ht]
\centering\leavevmode
\epsfxsize=150pt
\epsfbox{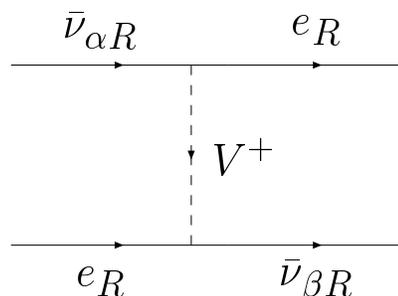}
\vglue -0.01cm
\caption{Vector contributions to anti-neutrinos FCNI in 331 model.  }
\label{fig3}
\end{figure}
From the above Eq.~(\ref{epsnu})  we can  explicitly calculate
$\epsilon^e_{\mu\tau}$ and $\epsilon^{'e}_{\mu\tau}=
\epsilon^e_{\tau\tau}-\epsilon^e_{\mu\mu}$ that enter Eq.~(\ref{motion})
for the  case of transitions $\nu_\mu \to \nu_\tau$, which can 
be relevant to atmospheric neutrinos:               
\begin{mathletters}
\label{epsnumutau}
\begin{eqnarray}
\epsilon^e_{\mu\tau}&=&\frac{1}{4\sqrt2G_FM^2_W}\;\left[
\left({\cal K}^\dagger_{LR} \right)_{1\mu}
\left( {\cal K}_{LR}\right)_{\tau1}\,x \right. \nonumber \\ &+&\left. 
 \left( {\cal K}^{\prime\dagger}_{LR}\right)_{1\mu}\left(
{\cal K}^\prime_{LR}\right)_{\tau1}\,y\right],
\label{epsmutau}
\end{eqnarray}
and
\begin{eqnarray}
\epsilon'^{e}_{\mu\tau}&=&\frac{1}{4\sqrt2G_FM^2_W}\;\left\{
\left[\left({\cal K}^\dagger_{LR} \right)_{1\tau}\left( {\cal K}_{LR}
\right)_{\tau1}\right.\right.
\nonumber \\ &-&\left.\left.
\left({\cal K}^\dagger_{LR} \right)_{1\mu}\left({\cal K}_{LR}
\right)_{\mu1} \right]x+ 
\left[ \left( {\cal K}^{\prime\dagger}_{LR}\right)_{1\tau}\left(
{\cal K}^\prime_{LR}\right)_{\tau1}\right.\right.
\nonumber \\ &-&\left.\left.\left( {\cal K}^{\prime\dagger}_{LR}\right)_{\mu1}
\left({\cal K}^\prime_{LR} \right)_{1\mu}\right]
\right\}.
\label{epspmutau}
\end{eqnarray}
\end{mathletters}

Similar for anti-neutrinos, where  ${\bar{\nu}}_\mu \to {\bar{\nu}}_\tau$
transitions are at play, we obtain from Eq.~(\ref{epsnubar}):
\begin{mathletters}
\label{epsbarnutau}
\begin{eqnarray}
& & {\bar{\epsilon}}^e_{\mu\tau}=\frac{1}{4\sqrt2G_FM^2_W}\;\left\{
({\cal K})_{1\mu}({\cal K}^\dagger)_{\tau1}z' \right.
\nonumber\\ &+& \left.
\left( {\cal K}_{LL}\right)_{1\mu}\left({\cal K}^\dagger_{LL}
 \right)_{\tau1}x'+
\left( {\cal K}^{\prime}_{LL}\right)_{1\mu}
\left( {\cal K}^{\prime\dagger}_{LL}\right)_{\tau1}y' \right\},
\label{epsbar}
\end{eqnarray}
and
\begin{eqnarray}
& & {\bar{\epsilon}}'^{e}_{\mu\tau}=\frac{1}{4\sqrt2G_FM^2_W}\;\left\{
\left[({\cal K})_{1\tau}({\cal K}^\dagger)_{\tau1}
-({\cal K})_{1\mu}({\cal K}^\dagger)_{\mu1}\right]
z'\right. \nonumber \\ &+&\left. 
[ \left({\cal K}_{LL}  \right)_{1\tau}\left({\cal K}^\dagger_{LL} 
\right)_{\tau 1}
- \left({\cal K}_{LL} \right)_{1\mu}\left( {\cal K}^\dagger_{LL}
\right)_{\mu1}]x'\right.
\nonumber \\ 
& & \mbox{}\left.+
\left[\left( {\cal K}^{\prime}_{LL}\right)_{\tau1}\left(
{\cal K}^{\prime\dagger}_{LL} \right)_{\tau1}
 -\left( {\cal K}^{\prime}_{LL}\right)_{1\mu}
\left({\cal K}^{\prime\dagger}_{LL} 
\right)_{\mu1}\right]y'
\right\}.
\label{epspbar}
\end{eqnarray}
\end{mathletters}

Notice that neutrinos and anti-neutrinos will not, in general, interact
in the same way through the boson fields. This is not a $CP$ violation effect
but rather only reflects the fact that these particles have different
interactions with matter in this theoretical scheme.
In fact, if all couplings and the vacuum 
expectation values are real, $CP$ is conserved in all the vertices of the 
model. The pure gauge boson interactions also conserve $CP$ if the $SU(3)_L$ 
gauge and $U(1)_N$ vector bosons $W^a\;\;\mbox{with}\,a=1,...8$,
and $B$, respectively transform as~\cite{cp3} 
{\small
\begin{eqnarray}
& (W^1_\mu,W^2_\mu,W^3_\mu,W^4_\mu,W^5_\mu,W^6_\mu,W^7_\mu,W^8_\mu,B_\mu)
\stackrel{CP}{\to} \\
& -(W^{1\mu},-W^{2\mu},W^{3\mu},-W^{4\mu},W^{5\mu},-W^{6\mu},
W^{7\mu},W^{8\mu},B^\mu).
\label{e12}
\end{eqnarray}
}
It means that the physical fields transform as 
{\small
\begin{eqnarray}
& (W^+_\mu, V^+_\mu,U^{++}_\mu,A_\mu,Z_\mu,Z^\prime_\mu)
\stackrel{CP}{\to}\\ & 
-(W^{\mu-}, -V^{\mu-},-U^{\mu--},A^\mu,Z^\mu,Z^{\prime\mu}).
\label{wvu}
\end{eqnarray}
}
where $W$ is the standard model gauge vector boson, 
$V$ is a gauge vector boson , $U^{++}$ is a doubly 
charged gauge boson and $Z^\mu$ and $Z^{\prime\mu}$ 
are neutral gauge vector bosons. 

In order to obtain the relevant FCNI parameters for solar neutrino analysis, 
{\it i.e.}, transitions involving $\nu_e$ and $\nu_\tau$, it is sufficient to 
change appropriately the indexes $\mu \to e$ in the 
Eqs.~(\ref{epsmutau}) and (\ref{epspmutau}).  
Notice also that in general $\vert\epsilon^f_{\alpha\beta}\vert\not=
\vert{\bar {\epsilon}}^f_{\alpha\beta}
\vert$, and similarly $\vert\epsilon'^f_{\alpha\beta}\vert\not=
\vert{\bar {\epsilon}}'^f_{\alpha\beta}
\vert$. In order to have
$\vert\epsilon^f_{\alpha\beta}\vert\approx\vert{\bar{\epsilon}}^f_{\alpha\beta}
\vert$ or $\vert\epsilon'^f_{\alpha\beta}\vert\approx
\vert{\bar{\epsilon}}'^f_{\alpha\beta}\vert$, we need
both the contributions of the gauge vector boson $V$  to be small and a fine
tuning among the Yukawa couplings and the mixing angles.

Another important remark concerning
the model is the following. In order to get to all charged leptons an 
appropriate mass only the sextet of scalars is necessary~\cite{331a}. 
However, unless we introduce an extra symmetry the 
triplet $\eta\sim({\bf1},{\bf3},0)$~\cite{331,331a}
couples also to the leptons. In this case we have FCNI at the tree level 
mediated by scalars. The mass matrix is diagonalized
by the biunitary transformation using the matrices $E_{L,R}$:
$\hat{M}^l=E^\dagger_L(Gv_s+G'v_\eta)E_R$ ($\hat{M}^l=
\mbox{diag}(m_e,m_\mu,m_\tau)$ and $v_s$ is the vacuum expectation
value (VEV) of one of neutral Higgs ( denoted $\sigma_2^0$ in the 
Eq.~(23) of Ref~\cite{331a}) contained in the
sextet~\cite{331a} and $v_\eta$ is the VEV of one of
triplets~\cite{331,331a}) but this transformation does not diagonalize
$Gv_s$ nor $G'v_\eta$ separately. Hence, 
the interactions in Eq.~(\ref{yu331}) are not suppressed by the lepton mass.
Since $v_s$ is the VEV  needed only in order to give mass to
the charged leptons it is not necessarily of the order of the hundred GeV but
may have a value of a few GeV. Hence at this stage the only constraints in 
the Yukawa parameters $G$ and $G'$ come from perturbation theory: 
$\vert G\vert^2/4\pi<1$, $\vert G'\vert^2/4\pi<1$. Since $G$ and $G'$ can be
arbitrary complex matrices, the  matrices defined in 
Eq.~(\ref{yu331}) are non-unitary matrices. 

We observe that in order to obtain $\epsilon^f_{\alpha\beta}$ and 
$\epsilon'^f_{\alpha\beta}$
appropriate to solve solar or atmospheric neutrinos anomalies
it is necessary that $x,y,x',y'$ be not too small (that is not the case
of grand-unified theories as discussed above). The charged scalars can not be 
so heavy.
In fact, even the constraints coming from the neutral kaon system do not
impose severe restriction to the mass of these scalar: 
$m_{h^+}\sim m_{h^0}\sim150$ GeV ($h^+$ and $h^0$ denote any of the  singly 
charged and neutral scalars) if at the same time we take into account
the mixing angles in the quark sector~\cite{dumm}. 
It means $x\sim y\sim 0.8$ already give
a value of $\epsilon^f_{\alpha\beta}\sim 0.1$ which is needed for solving the
atmospheric neutrino anomaly~\cite{rpatm}. A fine tuning of the
mixing angles keep $\epsilon'^f_{\alpha\beta}$ almost zero. However, it is 
still possible that $x,y,x',y'$ be larger than one because $h^+$ or $h^0$ 
contributing to the neutral kaons parameters are not necessarily the same
component than the scalars in Eqs.~(\ref{epsnumutau}). The
same analysis is valid for the anti-neutrino parameters in 
Eqs.~(\ref{epsbarnutau}). 

All scalars in Eq.~(\ref{yu331}) are still symmetry eigenstates. Hence,
they are linear combinations of the mass eigenstates. It means that some of 
their components couple to quarks and induce contribution to the neutral kaon
parameters. However, those contributions depend on mixing angles in the
$u$- and $d$- quarks mixing matrices so they do not impose constraints
on the couplings in Eq.~(\ref{yu331}). They will be constrained by some
phenomenological processes as discussed later in Sec.~\ref{sec:cons}. 

\section{Multi-Higgs doublets extension of the standard model}
\label{sec:mhesm}
Let us consider the standard model plus an arbitrary number of scalar doublets
$\Phi_i=(\phi^+_i , \phi^0_i)^T,\;i=1,2,...N\geq3$
In the lepton sector the Yukawa interactions are (with massless neutrinos)
\begin{equation}
-{\cal L}_Y^l=\sum_i\left( \bar{\nu}_{L}\,E^\dagger_L\Gamma^l_iE_R\,
l_{R}\phi^+_i +
\bar{l}_{L}\, E^\dagger _L\Gamma^l_{i}E_R\, l_{R}\,
\phi^0_i\right)
+H.c.,
\label{18}
\end{equation}
as in the previous model we have redefined the fields as in Eq.~(\ref{sme}). 
The mass matrix for the charged leptons 
$M^l=\sum_i(v_i/\sqrt2)\Gamma^l_i$ is diagonalized as follows:
\begin{equation}
E^\dagger_LM^lE_R=\hat{M}^l,
\label{19}
\end{equation}
with $\hat{M}^l={\rm diag}(m_e,m_\mu,m_\tau)$. 
Hence, the unitary matrices 
$E_{L,R}$ diagonalize $M^l$ but not each of the $\Gamma^l_i$ separately.
Although we have redefined the neutrino fields in the charged currents coupled
to the vector bosons $W$, the same is not 
possible in the interactions with $\phi^\pm_i$. 
Hence, even with massless neutrinos we
cannot avoid, in general, to have mixing in the charged current mediated by 
scalars in the lepton sector as well. Moreover, even if  neutrinos are
massless, there are flavor changing interactions mediated by neutral scalars
in the charged lepton sector.

The charged currents in Eq.~(\ref{18}) can be written in terms of the physical 
charged scalar ${\cal H}^+_i$ (defined as 
$\phi^+_i=\sum_iK_{ij}{\cal H}^+_j$) :
\begin{equation}
{\cal L}^{CC}=\sum_j\bar{\nu}_{L}\,{\cal V}_j\,l_{R}\, {\cal H} ^+_j+H.c.,
\label{18p}
\end{equation}
where 
\begin{equation}
({\cal V}_j)_{k,\alpha}=\sum_i\left(E^\dagger_L\Gamma^l_{i}E_R
K_{i j}\right)_{k \alpha},
\label{18pp}
\end{equation}
where $\alpha=e,\mu,\tau$ and  $k=1,2,3$. The matrices ${{\cal V}}_j$  
are not unitary matrices since $\Gamma^l$ are in general arbitrary complex 
matrices, with $\vert\Gamma\vert^2/4\pi<1$. 

In this case we have
\begin{equation}
\epsilon^e_{\alpha\beta}=\frac{1}{4\sqrt2G_FM^2_W}\;\sum_j
\left({\cal V}_j^\dagger\right)_{\alpha 1}\left( {\cal V}_j\right)_{\beta1}
\xi_j.
\label{epmh}
\end{equation}
Explicitly 
\begin{equation}
\epsilon^e_{\mu\tau}=\frac{1}{4\sqrt2G_FM^2_W}\;\sum_j
\left({\cal V}_j^\dagger \right)_{\mu1}\left( {\cal V}_j\right)_{\tau1}
\xi_j,
\label{ep}
\end{equation}

\begin{equation}
\epsilon'^{e}_{\mu\tau}\!\!=\frac{1}{4\sqrt2G_FM^2_W}\left[\sum_j
\left({\cal V}_j^\dagger \right)_{\tau 1}\left( {\cal V}_j\right)_{\tau1}
\!\!-\!\!
\left({\cal V}_j^\dagger \right)_{\mu 1}\left( {\cal V}_j\right)_{\mu1}\right]
\xi_j,
\label{epp}
\end{equation}
with $\xi_j=M^2_W/M^2_{{\cal{ H}}_j}$.
Note again that 
in order to obtain the relevant FCNI parameters for solar neutrino analysis 
it is sufficient to change appropriately the indexes $\mu \rightarrow e$ in 
the Eqs.~(\ref{ep}) and (\ref{epp}).
In this case $\epsilon_{\alpha\beta}=-{\bar{\epsilon}}_{\alpha\beta}$ 
(similarly for the 
$\epsilon'$s) and the constraints coming
from $\mu\to e\gamma$ and other processes are similar as for the previous 
model as it will be shown in the next section.

\section{Constraints on FCNI parameters}
\label{sec:cons}

In order to be sure that a realization of the oscillations induced by FCNI can 
happen in a quantitative way and a compatibility of data and related 
predictions be achieved we have finally to investigate the constraints to 
these FCNI  that arise from 
the charged lepton decay measurements and from the absence of
violation of the lepton number conservation laws~\cite{pdg}. 

The $\mu$-decay is measured with large precision and it is consistent with 
the standard  electroweak model predictions. This imposes severe constraints 
on exotic interactions involving neutrinos. Considering the 331 model, 
$H_2^-$ and $\eta_2^-$ mediate processes which should be summed coherently to 
the standard model contribution. The interference contribution of 
$H_2^-$ and $\eta_2^-$ with the standard model (SM) prediction, for the
the left-handed $\mu$-decay measurement, leads 
to~\cite{comment2}:


\begin{eqnarray}
 & \left| 
Re\left[ 
{({\cal K}'^\dagger_{LL})_{e2} ({\cal K}'_{LL})_{1\mu} \, y' } 
 + {({\cal K}^\dagger_{LL})_{e2} ({\cal K}_{LL})_{1\mu} \, x' } \right] \right|
\nonumber \\ 
&   <   (G_F/\sqrt{2}) M^2_W \Delta_\mu  , 
\label{mu-life}
\end{eqnarray}
where $\Delta_\mu$ is the combined experimental and theoretical 
error of $\mu$-decay. 
Following the recipe of Ref.~\cite{bottino} we compute the $G_{F}^{SM}$ 
value 

\begin{equation}
\label{eq:GF}
G_{F}^{SM} =\frac{ \pi\alpha(0)}{\sqrt{2}M_W^2\left( 1 - M_W^2 /M_Z^2 \right)
(1-\Delta r)} \;.
\end{equation}
In Eq. 
(\ref{eq:GF}) the fine
structure constant $\alpha(0)$ 
is very accurately known to be
1/137.036 and $\Delta r$ is the radiative SM correction and is found to 
be $\Delta r = .0349\mp.0019\pm.0007$~\cite{pdg}. However, in 
computing $G_{f}^{SM}$ the largest error comes from $M_{W}$
measurement with current value given as $M_{W} = 80.39\pm.06$ GeV~\cite{Kar}
and a much smaller error comes from $M_{Z} = 91.1867\pm.0020$ GeV~\cite{pdg}.

Using this and propagating the error of  $G_{F}^{SM}$ in the error of 
$\mu$-decay width we get that at $1\sigma$, 
$\Delta_\mu=8.6\times 10^{-3}$. The mainly
error came from the $G_{F}^{SM}$ uncertainty. Also in the case of $\tau$-decay
we need to take into account the $\tau$-lifetime measurement error and
$\tau$-mass measurement error, then
$\Delta_\tau=9.5\times 10^{-3}$ at $1\sigma$.


We can consider also the $\tau$-decay measurement 
\begin{eqnarray}
 & \left| 
Re\left[ 
{({\cal K}'^\dagger_{LL})_{e3} ({\cal K}'_{LL})_{1\tau} \, y' } 
 + {({\cal K}^\dagger_{LL})_{e3} ({\cal K}_{LL})_{1\tau} \, x' } \right] 
\right|
\nonumber \\ 
&   <  (G_F/\sqrt{2}) M^2_W \Delta_\tau  , 
\label{tau-life}
\end{eqnarray}
which constraint is obtained from 
Eq.~(\ref{mu-life}) substituting appropriately the sub-indexes and imposing  
$\Delta_\tau= 9.5\times 10^{-3}$ which leads to 
less stringent limits. 

We can rewrite, using the symmetric (anti-symmetric ) properties of 
${\cal K}_{LL}$ (${\cal K}'_{LL}$ ),  
\begin{eqnarray}
 & \left| 
Re\left[ 
{-({\cal K}'^\dagger_{LL})_{2e} ({\cal K}'_{LL})_{1\mu} \, y' } 
 + {({\cal K}^\dagger_{LL})_{2e} ({\cal K}_{LL})_{1\mu} \, x' } \right] \right|
\nonumber \\ 
&   <  (G_F/\sqrt{2}) M^2_W \Delta_\mu  , 
\label{mu-life1}
\end{eqnarray}
and 
\begin{eqnarray}
 & \left| 
Re\left[ 
{-({\cal K}'^\dagger_{LL})_{3e} ({\cal K}'_{LL})_{1\tau} \, y' } 
 + {({\cal K}^\dagger_{LL})_{3e} ({\cal K}_{LL})_{1\tau} \, x' } \right] \right|
\nonumber \\ 
&   <  (G_F/\sqrt{2}) M^2_W \Delta_\tau  , 
\label{tau-life1}
\end{eqnarray}
These constraints can give limits on the contribution of $H_2^-$ and
$\eta_2^-$ to the ${\bar{\epsilon}}^e_{\alpha\alpha}$ defined in
Eqs.~(\ref{epsnubar}). Respectively the $\mu$ decay and the 
$\tau$ decay can  give limits on 
${\bar{\epsilon}}^e_{22}$ and ${\bar{\epsilon}}^e_{33}$. 
Nevertheless, 
Eqs.~(\ref{mu-life1}) and ~(\ref{tau-life1}) can easily exhibits
cancellations among the contributions which can eliminate the constraints.


There are also other contributions of the charged-lepton decay mediated by 
$H_2^-$ and $\eta_2^-$ which do not sum coherently to the standard electroweak
process


\begin{eqnarray}
 & \sum_{\alpha,\beta}
\left| 
{({\cal K}'^\dagger_{LL})_{\alpha j} ({\cal K}'_{LL})_{i\beta} \, y' } 
 + {({\cal K}^\dagger_{LL})_{\alpha j} ({\cal K}_{LL})_{i\beta} \, x'
}  \right|^2
\nonumber \\ 
&   <  (G_F/\sqrt{2} M^2_W)^2 \Delta_j  , 
\label{mu-life3}
\end{eqnarray}
where $j=\mu,\tau $ and $i \neq  \alpha$ and $j \neq \beta$. 
These relevant constraint is to the  ${\bar{\epsilon}}^e_{\alpha\beta}$ defined in
Eq.~(\ref{epsnubar}) is given for $i=1$ and $\alpha=1$ in 
Eq.~(\ref{mu-life3}). Then the $H_2^-$ contribution to 
${\bar{\epsilon}}^e_{1\beta}$ is 

\begin{equation}
\left| {\bar{\epsilon}}^e_{1\beta}\right| _{H_2^-}  < \sqrt{\Delta_\beta}/8
\end{equation}
where $\beta=2,3$. The contribution of $\eta_2^-$ vanishes
when $\alpha=1$ because the anti-symmetric character of ${\cal K}_{LL}$.

Also we have the contributions of $H_2^-$ and $\eta_2^-$ that are
constrained

\begin{equation}
\left| {\bar{\epsilon}}^e_{23}\right|_{H_2^- \eta_2^-}  < 
\sqrt{\Delta_{\tau}}/8
\end{equation}

In the 331 model there are contributions to the right-handed charged-lepton 
decays mediated by $V_\mu^+$, $H_1^-$ and $\eta_1^-$ do not sum coherently to 
the standard process. The relevant constraint is

\begin{eqnarray}
& \sum_{\alpha,\beta}  \left|  {({\cal K}_{LR}^\dagger)_{i \beta} 
({\cal K}_{LR})_{\alpha j}  x }  + 
{({\cal K}_{LR}'^\dagger)_{i \beta} 
({\cal K}'_{LR})_{\alpha j} y }   \right|^2 +
\nonumber
\\ & 
{ \sum_{\alpha,\beta} 
\left| ({\cal K}^\dagger)_{\alpha j} ({\cal K})_{i \beta} \right|^2
{z'}^2  }  
  < (G_F^2/2)  M^4_W  {\Delta_\mu}
\label{mu-life_VNH}
\end{eqnarray}
with $\alpha$ and $\beta$ are related to the final neutrino states, 
$j$ and $i$ is related to the initial and final leptons respectively. 
If $j=3$ ($j=2$) implies $\tau$ ($\mu$)-decay.
Recall that although ${\cal K}$ is a unitary matrix, 
no symmetry relation exists in 
${\cal K}_{LR}$ or ${\cal K}'_{LR}$ couplings. 
Each of elements of the sum defined in Eq.~(\ref{mu-life_VNH}) also need 
to be constrained to be lower then the right side.

For example, the combination of product of matrices ${\cal
K}$ given in Eq.~(\ref{mu-life_VNH} did not appear in the definition
of $\bar{\epsilon}^e_{\alpha\beta}$ (Vide Eq.~(\ref{epsnubar})), but
some of matrices element can be constrained to be small. The element 
$({\cal K}^\dagger)_{\alpha j}$ with $j>1$, or 
$({\cal K})_{j\alpha}$ did not appear in the  Eq.~(\ref{epsnubar}),
but can be constrained in the $\tau\rightarrow \mu \gamma$ decay (see
below). Similar arguments apply for the 
${\cal K}_{LR}$ or ${\cal K}'_{LR}$ couplings.


Concerning the contributions to the charged lepton decays arising in the 
multi-Higgs models of Section~\ref{sec:mhesm}, there are no contributions to 
be coherently summed to the standard model contribution to the charged lepton 
decays. Therefore the relevant constraints coming from the charged lepton 
decay measurements can be written as

\begin{eqnarray}
&  \sum_{\alpha,\beta}  
{\left| \sum_j ({\cal V}_j)_{\alpha k} ({\cal V}^\dagger_j)_{l\beta} 
\xi_j \right|^2
 }
 & < (M^2_W G_F/ \sqrt{2})^2 \Delta_j,
\label{mu-life_higgs}
\end{eqnarray}

with $\alpha$ and $\beta$ related to the final neutrino states, 
$k$ and $l$ is related to the initial and final leptons respectively. 
Also each term of sum of Eq.~(\ref{mu-life_higgs}) is constrained to
be smaller then the right side of Eq.~(\ref{mu-life_higgs}).

Now, from Eq.~(\ref{epmh}) we can rewrite 

\begin{equation}
\epsilon^e_{\alpha\beta}=\frac{1}{4\sqrt2G_FM^2_W}\;\sum_j
\left({\cal V}_j^*\right)_{1 \alpha}\left( 
{\cal V}_j^{\dagger *}\right)_{1\beta}
\xi_j.
\label{epmh1}
\end{equation}

Then comparing Eq.~(\ref{epmh1}) and ~(\ref{mu-life_higgs}) we get

 \begin{equation}
\left| \epsilon^e_{\alpha\beta} \right|^2 < \Delta_{\alpha}/64
\label{epmh2}
\end{equation}
where $\alpha=2,3$ and any $\beta$. 

These is a stronger constraint on the multi-Higgs model. Only the element
 $\epsilon^e_{11}$ is not constrained.


Lepton number conservation tests~\cite{pdg} can generate strong limits to FCNI
parameters.
In 331 model and in the multi-Higgs model there are constraints coming 
from  $\mu\to e\gamma$, $\tau\to e\gamma$, $\tau\to \mu\gamma$ and 
$\mu\to eee$. 
In the multi-Higgs doublets model there are contribution to the
$\mu\to 3e$ decay mediated by the neutral Higgs flavor changing interactions
given in Eq.~(\ref{18}) it constrains the same matrix elements that appear
in the charged currents in Eq.~(\ref{18}) so, the mass of the neutral
Higgs bosons and the respective mixing angles. 

In 331, appear contributions to the $\mu\to e\gamma$ decay via the exchange
of a gauge vector boson  $V$  ($\mu\to V\nu^c_\alpha\to e\gamma$);
via exchange  of charged Higgs $H_1^-$ and $\eta_1^-$
($\mu\to H_1(\eta_1)\nu_\alpha\to e\gamma$), as well of  
$H_2^-$ and $\eta_2^-$ ; 
via exchange of doubly charged gauge boson $U^{--}$ (the 
interactions are not showed here)~\cite{331} and
via exchange of doubly charged Higgs $H^{++}_1$ ($H^{++}_2$).

The amplitude of the contribution of gauge vector boson $V$ 
is proportional to 
$\sum_\alpha{\cal K}_{2\alpha}{\cal K}^\dagger_{\alpha1}$ and vanish 
because ${\cal K}$ is a unitary matrix. 

The contribution of charged Higgs $H_1^-$ and $\eta_1^-$ can  be written as
\begin{eqnarray}
&   \left| \sum_\alpha { ({\cal K}^\dagger_{LR})_{1\alpha}}  
{ ({\cal K}_{LR})_{\alpha 2}
x } + 
         \sum_\alpha {  ({\cal K}'^\dagger_{LR})_{1\alpha}
({\cal K}'_{LR})_{\alpha 2}  y }  
 \right|^2 
\nonumber 
\\ & < (48\pi/\alpha(0))  M^4_W BR(\mu\to e\gamma) 
\label{mu-egamma_1}
\end{eqnarray}
where $BR(\mu\to e\gamma)< 1.2\times 10^{-11}$~\cite{mueg} is the branching 
ratio of the lepton violating process $\mu\to e\gamma$. Note, however, that 
the coupling $({\cal K}_{LR})_{\alpha 2}$ and $({\cal K}'_{LR})_{\alpha 2}$ 
do not appear in 
Eqs.~(\ref{epsnu}) and (\ref{epsnubar}) and therefore the limit above does 
not imply direct  constraint on ${\epsilon}^e_{\alpha\beta}$ and 
$\bar{\epsilon}^e_{\alpha\beta}$.

The same
happens with the contributions of $H_2^-$ and $\eta_2^-$.
Only the sum of some matrix elements is constrained:
\begin{eqnarray}
& \left| \sum_\alpha { { ({\cal K}^\dagger_{LL})_{\alpha 2} 
({\cal K}_{LL})_{1\alpha} x' }}
       + \sum_\alpha { { ({\cal K}'^\dagger_{LL})_{\alpha 2} 
({\cal K}'_{LL})_{1\alpha} y' }} \right|^2 
\nonumber
\\ & < (48\pi/\alpha(0))  M^4_W BR(\mu\to e\gamma) 
\label{mu-egamma_2}.
\end{eqnarray}
Even considering that the matrix ${\cal K}_{LL}$ (${\cal K}'_{LL}$) is 
symmetric (antisymmetric), their combination appearing in 
Eq.~(\ref{mu-egamma_2}) does not contribute to the definition of 
$\bar{\epsilon}^e_{\alpha\beta}$. Furthermore
the constraint involves a coherent sum in which we can find cancellations 
making it weaker.

Finally there is a contribution of the doubly charged vector boson $U^{--}$,
$\mu\to U^{--}l^+\to e\gamma$; since the mass of the charged leptons
are quite different, the amplitude only constrains
$|\sum_\alpha {\cal K}_{\mu\alpha}{\cal K}^\dagger_{\alpha e}m^2_\alpha/
m^2_{U^{--}}|^2$, with
$m_\alpha=m_e,m_\mu,m_\tau$. Since the $m_\tau$ dominates, the decay 
constrains only ${\cal K}^\dagger_{\mu3}{\cal K}_{3 e}$. These contributions 
should be coherent summed to the $H_2^{--}$ and to the $H_1^{--}$. The former
have the couplings ${\cal K}_{RR}$ (See Eq.~(\ref{yu331})) that did not appear
in the Eqs~(\ref{epsnu}) and (\ref{epsnubar}) and the later ${\cal K}_{LL}$
that appear in those equations, but instead of $H_2^-$ mass in the 
denominator of 
the Eqs~(\ref{epsnu}) and (\ref{epsnubar}) you have the $H_2^{--}$ mass
in the denominator.
You can choose to put the masses of $H_2^{--}$ field large enough to suppress
the constraints on ${\cal K}_{LL}$.

The $\mu\to 3e$ decay arise by the interaction
with the doubly charged Higgs 
$H^{++}_1$ ($H^{++}_2$) and by the interaction with the doubly charged
vector boson $U^{++}$. The Higgses $H^{++}_1$ ($H^{++}_2$) constrain
the matrix ${\cal K}_{LL}$ (${\cal K}_{RR}$) and 
doubly charged
vector boson $U^{++}$ which constrains the ${\cal K}$ mixing matrix.
The matrix elements involved
are ${\cal K}^\dagger_{\mu1}$ and ${\cal K}_{e1}$ and only the first one 
appears in Eq.~(\ref{epsbar}). Note, however, that the doubly charged vector 
and scalar bosons do not contribute to $\bar{\epsilon}^e_{\alpha\beta}$. Any 
constraint can be evaded by imposing that  their masses are sufficiently large.

At first sight the process $\mu\nu_e\to e\nu_\mu$ would impose constraints
on $\left( {\cal K}_{LR}\right)_{1\mu}$ and  
$\left( {\cal K}^\prime_{LR}\right)_{1\mu}$ and other parameters involving the 
$\mu$. However, this process has a cross section which is $0.90\pm0.20$ times 
the prediction of the $V-A$ theory~\cite{gargamelle} 
(or $0.98\pm0.18$~\cite{charm}) which does not imply therefore strong 
constraints on the related couplings. The doubly charged scalars 
$H^{++}_{1,2}$
may contribute to the muonium to anti-muonium conversion with the same
strength as the doubly charged vector bilepton. It depends on the values
of the parameters such as vacuum expectation values of  the 
model~\cite{muonprd}.

Finally we mention that the FCNI parameters can be constrained 
using only charged leptons decays and some violating lepton number processes.
We get that the contributions of charged Higgs scalars $H_2^-$ and $\eta_2^-$ 
to $\bar{\epsilon}^e_{1\beta}$, with $\beta=2,3$ and 
$\bar{\epsilon}^e_{23}$ are severely limited. Also some combinations of 
couplings are constrained to be very small like Eqs~(\ref{mu-life1}),
(\ref{tau-life1}), (\ref{mu-egamma_1}) and (\ref{mu-egamma_1}) 
although have no direct comparison with the relevant 
couplings of $\epsilon^e_{\alpha\beta}$ and 
$\bar{\epsilon}^e_{\alpha\beta}$. For the other couplings like
${\cal K}_{LR}$,  ${\cal K}'_{LR}$ and ${\cal K}$ we have a compromise 
between different processes but no direct limit can be obtained. 

For the case of multi-Higgs, the situation is  different. Only the charged
lepton decays impose a stronger limits on the  $\epsilon^e_{\alpha\beta}$ and 
$\bar{\epsilon}^e_{\alpha\beta}$. Unless you enlarge the model, {\it e. g.}
adding a set of singly charged Higgs which can made less drastic the 
constraints, we have no hope to get FCNI parameters large enough to get a
relevant oscillation.

Also is worth to mention that the $\epsilon^e_{\alpha\beta}$ and 
$\bar{\epsilon}^e_{\alpha\beta}$ parameters are essentially different,
then even in the case of massless neutrinos, the neutrinos and
anti-neutrinos have different evolution in the matter. This is not the
case in the $R$-parity violating supersymmetric models.

\section{Conclusions}
\label{sec:con}

Both models discussed above can generate FCNI with strictly massless neutrinos
with strength sufficient to induce appreciable effects both for solar 
as well as atmospheric neutrinos.
These effects can either be confirmed or ruled out as
the collected data sample increases and the solar and atmospheric
results become more precise.

If confirmed,  it is necessary to know what is  the actual theoretical
framework that can account for these type of FCNI. In fact, there are
very few models in the literature that introduced FCNI while keeping
neutrinos massless. If, on the other hand, future data points
undoubtedly in the direction of conventional neutrino oscillation (in vacuum 
or resonantly enhanced by matter), FCNI
present in these models will have to be suppressed and these models
will either be ruled out or their FCNI couplings severely restricted by
data.

It is worth to mention that a very interesting intermediate scenario occurs 
when oscillations induced by   $\bar{\epsilon}^f_{\alpha\beta}$ and 
$\bar{\epsilon}'^f_{\alpha\beta}$ 
co-exist with conventional neutrino  oscillations. It was pointed out that  
in this situation  simultaneously solution to the solar, atmospheric and LSND 
observations~\cite{lsnd} can be obtained in a context of  only three neutrino 
families~\cite{mr98}, dispensing therefore the introduction of a fourth light 
(electroweak singlet) neutrino. 

Constraints from several processes previously analyzed can generate limits 
to the FCNI parameters 
${\epsilon}^e_{\alpha\beta}$ and $\bar{\epsilon}^e_{\alpha\beta}$. We
see above that the 331 model survives like a model able to generate
FCNI parameters compatible with the constraints and a viable model of
FCNI and still keeping the neutrinos massless. This is a 
different situation from what happens in $R$-parity broken supersymmetric 
models where the combination of matrices entering in the definition of the 
FCNI parameters appear also contributing to constrained lepton violating 
processes. It is also different from what happens in the Zee model. There the 
unique anti-symmetric coupling makes it impossible  to avoid the constraints.

\begin{figure}[ht]
\centering\leavevmode
\epsfxsize=150pt
\epsfbox{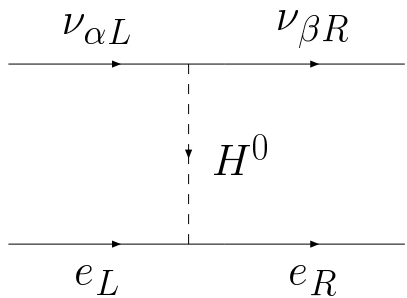 }
\vglue -0.01cm
\caption{Contribution if right-handed neutrinos are added.  }
\label{fig4}
\end{figure}

So far, neutrinos remain massless. In both models we can get massive
neutrinos simply by adding the right-handed components and the neutrino masses
are arbitrary. In the 331 model we have still the possibility that one
of the neutral component of the scalar sextet~\cite{331a} ( denoted 
$\sigma_1^0$ in the Eq.~(23) of Ref~\cite{331a}) get a non-vanish VEV, here 
denoted by $v_1$, giving to the neutrinos a majorana mass term 
\begin{equation}
\left(\begin{array}{cc}
Gv_1 & \frac{1}{2}G'v_\eta\\
\frac{1}{2}G'v_\eta & M^R
\end{array}
\right)
\label{nmm}
\end{equation}
where $M^R$ is a possible Majorana mass term for the right-handed 
singlets. The neutrino masses are still arbitrary but there are FCNI
in the scalar sector. It means that there are Yukawa couplings that
are not proportional to the neutrino masses.
In both models we can also add several neutral or doubly charged scalar
singlets only to get calculable neutrino masses~\cite{zee,cl}.
Or, it is possible to get calculable neutrino masses breaking explicitly the
total lepton number in the scalar potential~\cite{fkl}.
In all these possibilities the neutrino masses do not depend on the 
parameters entering in the $\epsilon$ and $\epsilon^\prime$. One
example of new contribution to the FCNI effect, when the right-handed
neutrino is introduced,  is showed in  Fig.~\ref{fig4}.

Hence, we see that there are no strong constraints on all the
parameters of the 331 model. And from this we hope that show a viable
model for flavor changing neutrino interactions with strictly massless
neutrinos that survive the constraints from charged lepton decays and some
violating processes like $\mu\rightarrow e\gamma$, $\mu\rightarrow
eee$. 

\acknowledgments 
This work was supported by Funda\c{c}\~ao de Amparo \`a Pesquisa
do Estado de S\~ao Paulo (FAPESP), Conselho Nacional de 
Ci\^encia e Tecnologia (CNPq) and by Programa de Apoio a
N\'ucleos de Excel\^encia (PRONEX). 
One of us (VP) would like to thank useful discussions with G. Castelo Branco.


\begin{references}
\bibitem{comment2} Note finally that stronger constraints can be obtained by 
considering the angular distribution in the charged lepton decay by 
introducing the Michel parameters as in
E. D. Carlson and P.H. Frampton, Phys. Lett. {\bf B283}, 123 (1992).
\end{references}
\end{document}